\documentclass[journal=jacsat,manuscript=article]{achemso}

\usepackage{chemformula} 
\usepackage[T1]{fontenc} 
\usepackage{lipsum}
\usepackage{siunitx}
\usepackage{graphicx}
\graphicspath{{./}{./Fig/}{./figures/}}
\usepackage[caption=false]{subfig}
\usepackage{amsmath}
\usepackage{verbatim}
\usepackage{float}
\usepackage{tabularx}
\usepackage{threeparttable}
\usepackage{dcolumn}
\usepackage{bm}
\usepackage{color,epsfig,multirow}
\usepackage{array}
\usepackage{hyperref}
\usepackage{algpseudocode}
\usepackage{algorithmicx,enumerate,algorithm}
\usepackage{booktabs}
\usepackage{threeparttable}
\usepackage{dcolumn}
\usepackage{geometry}

\definecolor{yscol}{HTML}{6622AA}

\usepackage{xcolor} 
\definecolor{tcr}{named}{red}        
\definecolor{tcb}{named}{blue}

\author{Xiaoqing Liu}
\affiliation{School of Mathematical Sciences, MOE-LSC and CMA-Shanghai, Shanghai Jiao Tong University, Shanghai 200240, China}
\alsoaffiliation{Shanghai Jiao Tong University–Chongqing Institute of Artificial Intelligence, Chongqing 401329, China}

\author{Xinyu Yu}
\affiliation{Future Battery Research Center, Global Institute of Future Technology, Shanghai Jiao Tong University, Shanghai 200240, China}
\alsoaffiliation{Global College, Shanghai Jiao Tong University, Shanghai 200240, China}

\author{Yangshuai Wang}
\affiliation{Department of Mathematics, National University of Singapore, 10 Lower Kent Ridge Road, Singapore}

\author{Zhe-Tao Sun}
\affiliation{Future Battery Research Center, Global Institute of Future Technology, Shanghai Jiao Tong University, Shanghai 200240, China}
\alsoaffiliation{Global College, Shanghai Jiao Tong University, Shanghai 200240, China}

\author{Zedong Luo}
\affiliation{Shanghai Jiao Tong University–Chongqing Institute of Artificial Intelligence, Chongqing 401329, China}

\author{Kehan Zeng}
\affiliation{Shanghai Jiao Tong University–Chongqing Institute of Artificial Intelligence, Chongqing 401329, China}

\author{Teng Zhao}
\email{zhaoteng_sjtu@sjtu.edu.cn}
\affiliation{SOG AI-Technology Co. Ltd., Shanghai 200240, China}

\author{Shou-Hang Bo}
\email{shouhang.bo@sjtu.edu.cn}
\affiliation{Future Battery Research Center, Global Institute of Future Technology, Shanghai Jiao Tong University, Shanghai 200240, China}
\alsoaffiliation{State Key Laboratory of Synergistic Chem-Bio Synthesis, School of Chemistry and Chemical Engineering, Shanghai Jiao Tong University, Shanghai 200240, China}

\author{Zhenli Xu}
\email{xuzl@sjtu.edu.cn}
\affiliation{School of Mathematical Sciences, MOE-LSC and CMA-Shanghai, Shanghai Jiao Tong University, Shanghai 200240, China}

\footnotetext[1]{X. Liu, X. Yu and Y. Wang contribute equally to this work.}
\footnotetext[2]{Corresponding authors: T. Zhao, S.-H. Bo and Z. Xu.}

\title[Fine-tuning and stress-testing MLIPs for battery interfaces]{
FIRE: Fine-Tuning and Stress-Testing Universal Interatomic Potentials for Solid–Solid Battery Interfaces
}

\abbreviations{IR,NMR,UV}

\begin{document}

\newpage
\vspace{1cm}
\begin{abstract} 

Fine-tuning foundation models is the standard route to machine-learning interatomic potentials, yet low benchmark errors do not show that a model can be trusted at a reconstructing battery interface. Here we present FIRE, which combines representative sampling with replay-augmented fine-tuning of MACE-MPA-0, and test it on six battery interface and electrolyte systems. FIRE reaches 0.1--1.0 meV/atom and 8--54 meV/\AA{} with 2--12\% of the data of the three largest datasets, and reproduces measured conductivities and elastic moduli. Deployed on a new 2712-atom Na$|$\ch{Na3SbS4}$|$Na interface, the model collapses within 50~fs. DFT dimer curves show that fine-tuning erodes the short-range repulsion inherited from pretraining, in all seven pairs across five systems, unnoticed by 57 bulk simulations. A model-only stress test exposes the defect before deployment, and 260 targeted configurations repair it without changing benchmark accuracy. The repaired model still fails at the interface: a static stress test is necessary but not sufficient.

\end{abstract}

\section{Introduction}
Solid--solid interfaces play a central role in governing the electro-chemo-mechanical behavior of battery materials, especially in solid-state batteries~\cite{SS_interface_Joule, PRL_Sun, AEL_Chen}. Unlike solid--liquid interfaces in conventional batteries, solid--solid interfaces exhibit intrinsic structural and chemical complexity, which poses formidable challenges for atomistic modeling. These challenges arise from several aspects. First, the configurational diversity induced by defects and dislocations substantially increases the difficulty of \emph{sampling} representative structures. Second, the heterogeneous local environments and complex bonding patterns can render conventional theoretical descriptions insufficient. In particular, battery interfaces often involve multiple crystalline phases, giving rise to lattice mismatch and interfacial strain. The coexistence of diverse chemical bonds further demands models with strong generalization across chemical space. Finally, multiphysics coupling, namely electro-chemo-mechanical coupling at solid--solid interfaces, makes it difficult for single-scale approaches to fully capture interfacial behavior~\cite{JPCL_SUN}.

Recent advances in machine-learning interatomic potentials (MLIPs)~\cite{behler2007generalized, wang2018deepmd, deringer2020modelling, euchner2022atomistic, olsson2022atomic, bartok2010gaussian, DrautzACE, schunet_2018, batatia2022mace, CACE_2024, ji2025machine} have opened the door to quantum-accurate simulations at scales far beyond the reach of first-principles methods. However, constructing robust MLIPs via from-scratch training remains highly data-intensive, often demanding a huge number of costly quantum calculations. This bottleneck is particularly severe for interfaces, where structural diversity and complex interactions generate vast configurational landscapes.
In parallel, fueled by large-scale pretraining and the rapid growth of materials databases such as the Materials Project~\cite{jain2013commentary}, universal machine-learning interatomic potentials (U-MLIPs) have emerged as general-purpose force fields, including MACE-MP-0~\cite{batatia2023foundation}, CHGNet~\cite{deng2023chgnet}, EquiformerV2~\cite{liao2023equiformerv2}, MatterSim~\cite{yang2024mattersim}, and DPA~\cite{zhangPretrainingAttentionbasedDeep2024a}, among others~\cite{neumann2024orb, xie2024gptff, park2024scalable, merchant2023scaling, nomura2025allegro}. By pretraining on chemically diverse datasets, these models can capture a broad spectrum of atomic interactions and exhibit strong generalization across material classes. U-MLIPs have been increasingly applied to battery materials~\cite{kim2024probing, staacke2021role, ju2025application, kim2022flexible, hajibabaei2021universal, wang2024accelerating, diddens2022modeling, wang2023strategies, zheng2024computational, maevskiy2025predicting, zhong2025machine}, offering a promising route toward scalable and transferable atomistic modeling of battery materials and interfaces. Nevertheless, interface problems remain fundamentally limited by \emph{data quality}: without representative interfacial configurations, even strong foundation models may miss rare yet mechanistically important local environments. Therefore, an efficient and robust \emph{sampling} pipeline (capable of generating, organizing, and selecting diverse interface structures at low cost) is essential for achieving quantum-level accuracy in interfacial modeling.

Here we address both halves of this problem. We first present Fine-tuning with Integrated Replay and Efficiency (FIRE), a data-efficient workflow for adapting a universal potential to a target interface: a lightweight pre-adaptation stage that enables stable molecular-dynamics (MD) generation, descriptor-based dimensionality reduction of the candidate pool, and diversity-driven selection of a compact fine-tuning set, followed by replay-augmented fine-tuning that retains the pretrained representation~\cite{li2023task}. Across six battery interface and electrolyte systems this yields models whose errors, ionic conductivities and elastic moduli agree closely with reference calculations and experiments. We then ask what these validations do and do not establish. Deploying the \ch{Na/Na3SbS4} model on a freshly built electrode--electrolyte interface exposes an unphysical collapse that none of the conventional tests anticipated; density-functional dimer calculations trace it to a loss of short-range repulsion introduced by fine-tuning and shared by the other systems; a model-only stress test detects the defect before deployment; targeted short-range sampling repairs it within the same workflow; and the repaired model still fails at the reconstructing interface. The outcome is a workflow that is accurate and data-efficient where its validations apply, together with a clear statement of where they stop.
FIRE is broadly applicable beyond battery systems, offering a practical solution to data scarcity and enabling deeper understanding of multiphysics coupling at complex interfaces.

\begin{figure*}
\centering
\includegraphics[width=15 cm]{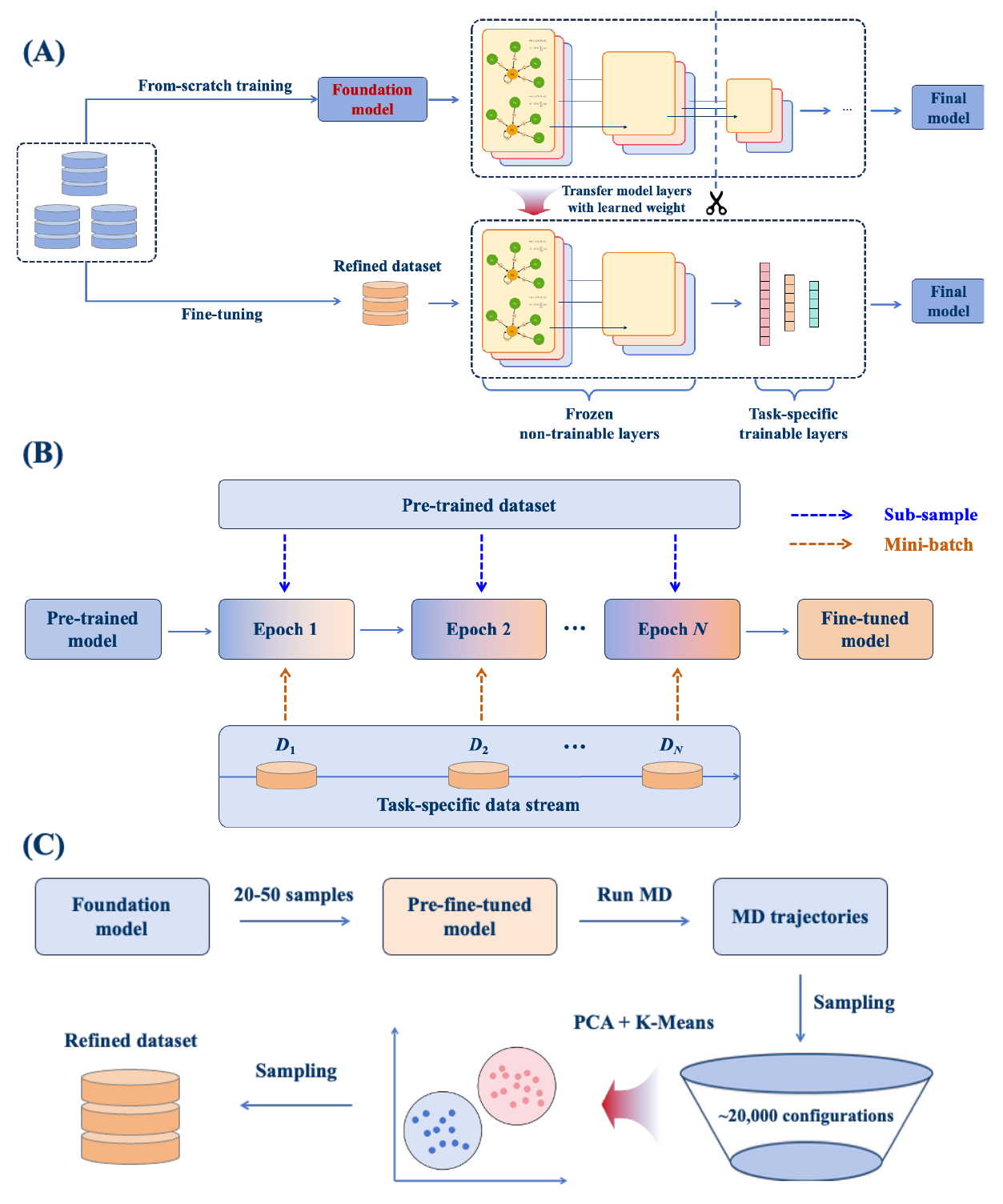}
\caption{Overview of the FIRE framework: (A) comparison between from-scratch training and fine-tuning paradigms; (B) replay-augmented continual fine-tuning strategy that leverages both pretraining and task-specific datasets; (C) efficient sampling workflow, including a pre-fine-tuned model for robust MD generation, dimensionality reduction, and diversity-based selection of representative configurations.}
\label{fig:workflow}
\end{figure*}

\section{Results and Discussion}
The schematic diagram on the key components of FIRE is illustrated in Figure~\ref{fig:workflow}. These components are designed to build accurate MLIPs with significantly reduced computational cost. As shown in Fig.~\ref{fig:workflow} (A), we adopt a task-specific fine-tuning paradigm instead of the from-scratch training, so that   the expressive architecture of a universal foundation model is retained with task-relevant parameters being updated selectively. This strategy preserves transferability and inductive biases inherited from large-scale pretraining, and significantly lowers data and training time requirements.
To further improve transferability and stability, we incorporate standard continual-learning practices during adaptation (Fig.~\ref{fig:workflow} (B)); we note that our primary novelty lies not in the fine-tuning mechanics themselves, but in enabling effective adaptation through high-quality data.

Crucially, we design an efficient and modular \emph{sampling} protocol (Fig.~\ref{fig:workflow} (C)) to construct high-quality training datasets for solid--solid interfaces. The workflow starts from a lightweight pre-adaptation step (trained on only 20--50 perturbed configurations) to obtain a sufficiently stable model for MD-based structure generation, which is often the practical bottleneck for complex interfacial systems. We then run MD to generate diverse atomic trajectories under realistic dynamics, and organize the resulting candidate pool ($\sim$20,000 configurations) using PCA-assisted dimensionality reduction and K-means clustering on SOAP descriptors~\cite{Bartok13Soap, Himanen20DScribe}. By selecting cluster representatives, we obtain a compact yet informative dataset with broad structural coverage and reduced redundancy. This sampling strategy directly improves data efficiency and provides an AI-ready dataset for downstream training~\cite{aiready_NSR}. Full details on data preparation, including first-principles labeling and the sampling workflow, are provided in Sections~S.1 and S.2 of the \textit{Supplementary Information} (SI).

\begin{figure*}
\centering
\includegraphics[width=16 cm]{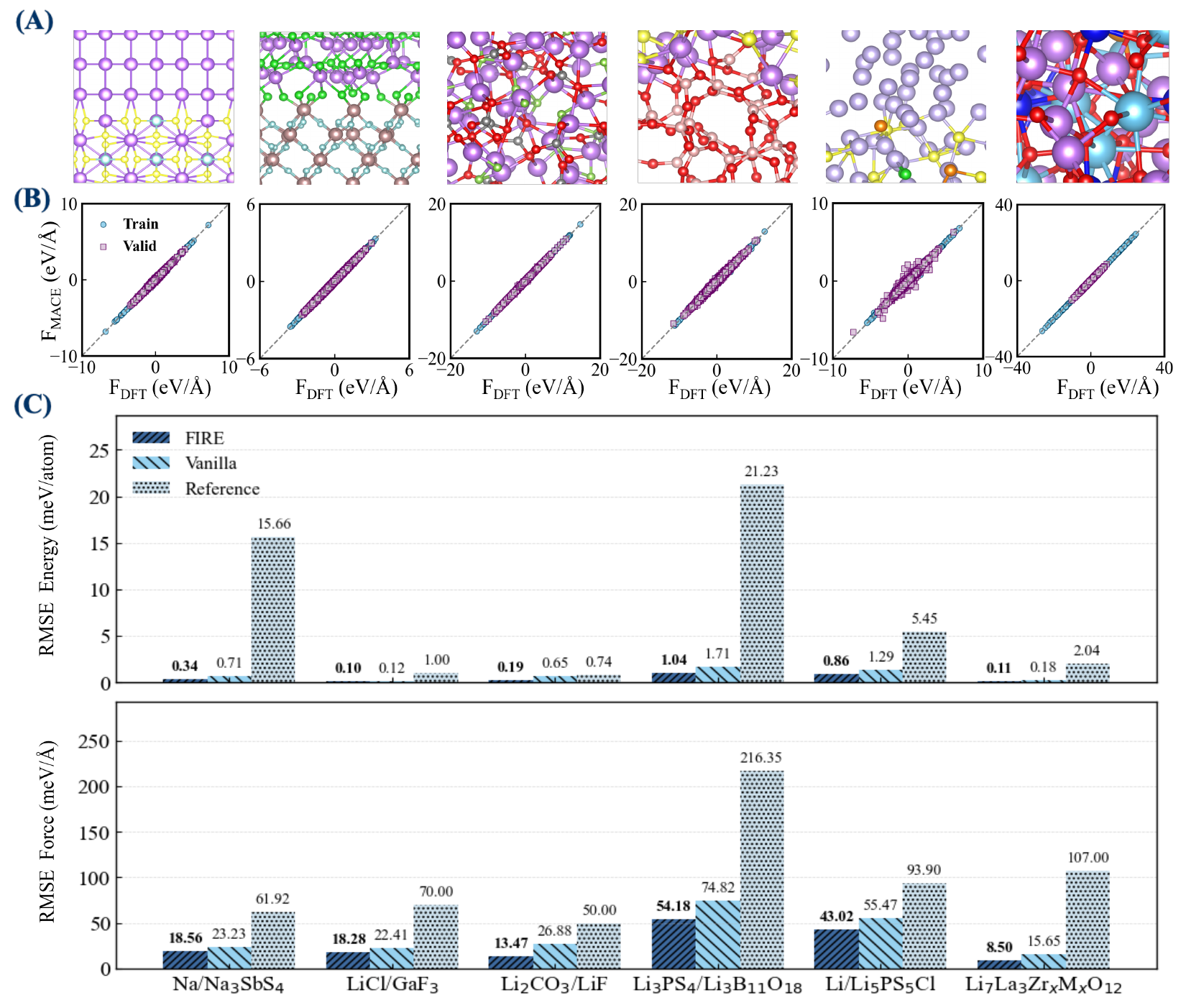}
\caption{Model evaluation of the FIRE approach: (A) atomic configurations of the six battery materials studied in this work, from left to right: \ch{Na/Na3SbS4}, \ch{LiCl/GaF3}, \ch{Li2CO3/LiF}, \ch{Li3PS4/Li3B11O18}, \ch{Li/Li6PS5Cl}, and \ch{Li_{7}La_{3}Zr_{2-x}M_{x}O_{12}} ($M = \mathrm{Ta}, \mathrm{Nb}$); (B) the corresponding force predicted by fine-tuned model with respect to that of DFT calculation; (C) model evaluation via RMSE for energy and force, respectively. The \ch{Na/Na3SbS4} entry is the 2000-frame run of Fig.~\ref{fig:efficiency}; the model deployed in Fig.~\ref{fig:application} is the later 2700-frame run, whose errors are 0.5 meV/atom and 19 meV/\AA{} (Section~S.6). For comparison, results collected from literature and vanilla fine-tuning are shown.}
\label{fig:accuracy}
\end{figure*}

\begin{figure*}
\centering
\includegraphics[width=16 cm]{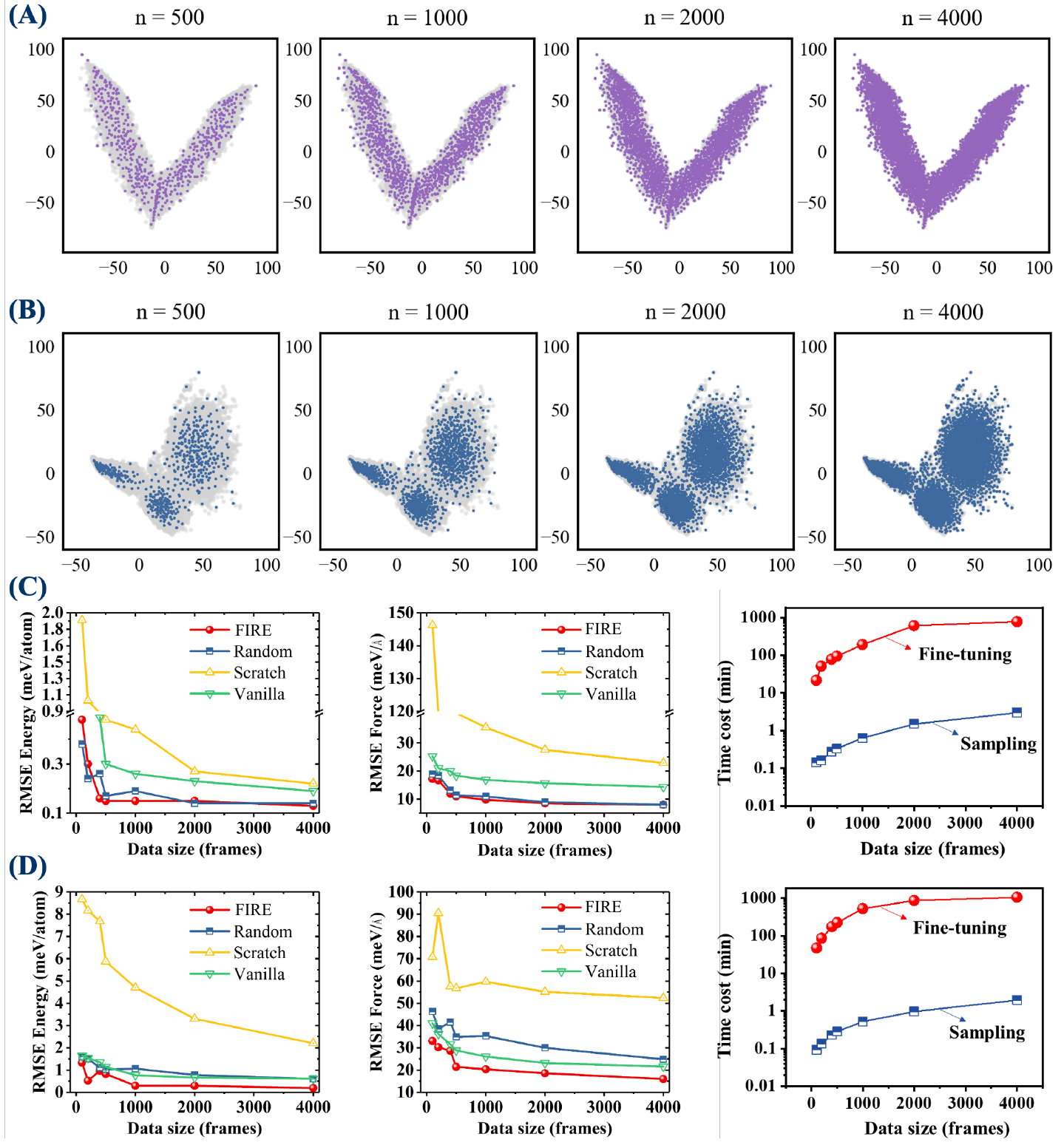}
\caption{Data efficiency of the FIRE approach: (A–B) PCA projections of sampled configurations for \ch{Na/Na3SbS4} and \ch{Li_{7}La_{3}Zr_{2-x}M_{x}O_{12}} ($M = \mathrm{Ta}, \mathrm{Nb}$), showing broader structural coverage with increasing dataset size; (C–D) data-efficiency benchmarks: replay-augmented fine-tuning achieves the lowest energy and force errors with reduced computational cost, outperforming vanilla fine-tuning, random sampling, and from-scratch training.
}
\label{fig:efficiency}
\end{figure*}

To evaluate FIRE, we select six typical solid-solid interfaces in batteries that capture chemical and structural complexity of practical systems. These include electrode–electrolyte (\ch{Na/Na3SbS4}, 
\ch{Li/Li6PS5Cl})~\cite{An2025LiNucleation, leeDisorderDependentLiDiffusion2024}, coating in composite cathodes (\ch{Li3PS4/Li3B11O18})~\cite{wang2023nature}, and solid electrolyte interphase at anodes (\ch{Li2CO3/LiF})~\cite{AISSquare_Li2CO3_LiF_2025} interfaces, as well as (electro)chemically disordered solid electrolyte (\ch{Li_{7}La_{3}Zr_{2-x}M_{x}O_{12}} ($M=\mathrm{Ta}, \mathrm{Nb}$), \ch{LiCl/GaF3})~\cite{you2024_pca, kim2024_learning}. The systems considered in this work capture key interfacial behaviors in solid-state batteries, providing a representative benchmark for evaluating the fine-tuning strategies. Four of the six are electrode--electrolyte or electrolyte--electrolyte interface models; the two disordered electrolytes test the workflow beyond planar interfaces. The snapshots of atomic configurations of these materials are present in Fig.~\ref{fig:accuracy} (A).
We adopt MACE-MPA-0~\cite{batatia2023foundation} as the backbone U-MLIP due to its strong reported performance and scalability, but emphasize that FIRE is conceptually general and applicable to other universal interatomic potentials. 
Fig.~\ref{fig:accuracy}~(B,C) show the results of energy and force calculations for different models.  
The force predictions of the fine-tuned models against DFT calculations (Fig.~\ref{fig:accuracy} (B)) present well agreements for all systems, indicating that FIRE preserves quantum-level accuracy even in heterogeneous environments.
The accuracy in terms of energy and force RMSEs is measured and shown in Fig.~\ref{fig:accuracy} (C). Compared with the from-scratch training, vanilla fine-tuning (viz. without replay mechanism), and previously reported results, the FIRE achieves the lowest errors, with energy RMSEs of 0.1--1.0 meV/atom (below 1 meV/atom for five of the six systems) and force RMSEs of 8--54 meV/Å (below 20 meV/Å for four of them) across diverse solid–solid interfaces. Current MLIPs for battery materials typically report energy errors exceeding 3 meV/atom and force errors surpassing 100 meV/Å (RMSE or MAE)~\cite{renVisualizingSEIFormation2024, ouAtomisticModelingBulk2024, laiUnravelingChargeEffects2025, Chandrappa2022,liAtomisticStudyReactivity2025, jangMechanisticInsightsSuperionic2025, you2024_pca, leeDisorderDependentLiDiffusion2024}. These literature values were obtained on different datasets and with different error measures and are quoted for orientation only; the controlled comparison is the one against our own baselines on identical data (Fig.~\ref{fig:efficiency} and Section~S.6), where at 4000 frames FIRE lowers the force RMSE by a factor of 1.6 relative to vanilla fine-tuning and 3.3 relative to from-scratch training. Hence, FIRE establishes a practical paradigm for constructing significantly more accurate MLIPs. Details about frameworks including loss curves and accuracies reported in the existed works are summarized in Fig.~S1 and Table~S2 in SI, respectively.

The predictive performance and computational efficiency of FIRE are assessed across different dataset sizes. Fig.~\ref{fig:efficiency} (A,B) shows PCA projections of sampled configurations for two representative systems: \ch{Na/Na3SbS4} (top) and \ch{Li_{7}La_{{3}}Zr_{2-x}M_{x}O_{12}} ($M = \mathrm{Ta}, \mathrm{Nb}$) (bottom). As the dataset size increases from 500 to 4000 frames, the configurational manifolds become increasingly well resolved, indicating enhanced structural diversity and more comprehensive coverage of the relevant configurational space. This directly evidences the effectiveness of our sampling protocol in capturing diverse interfacial environments. Details of the sampling procedure are provided in Section~S.1 in the SI. The accuracy and cost results in Fig.~\ref{fig:efficiency} (C,D) further show that FIRE achieves the lowest force RMSE at every dataset size and the lowest energy RMSE from 200 frames upwards, consistently outperforming baselines such as random sampling and from-scratch training. At the smallest budget of 100 frames the energy RMSEs of FIRE and of vanilla fine-tuning differ by less than the run-to-run scatter, so the energy advantage is established only once a few hundred configurations are labelled. Importantly, the improved accuracy is achieved with substantially fewer labeled structures because the sampling stage reduces redundancy and prioritizes representative configurations. The from-scratch, random-sampling and FIRE runs share one training protocol (300 epochs, batch size 8, energy-to-force loss weights of 1:10, identical optimizer, learning rate and stochastic weight averaging), so Fig.~\ref{fig:efficiency}(C,D) compares models at a matched training budget rather than separately converged baselines; the vanilla fine-tuning baseline instead used weights of 10:1 and a batch size of 4, which favours its energy error at the expense of its force error (Section~S.6). Each point is a single training run unless stated otherwise: repeated FIRE runs scatter by 1--2 meV/\AA{} in force RMSE, whereas from-scratch training at 100 frames scatters by 19 meV/\AA{} over ten seeds, which exceeds several of the differences between neighbouring dataset sizes. Overall, these results highlight sampling as the key lever for data efficiency in interfacial MLIPs, enabling competitive accuracy at substantially reduced data and computational cost.

\begin{figure*}
\centering
\includegraphics[width=16 cm]{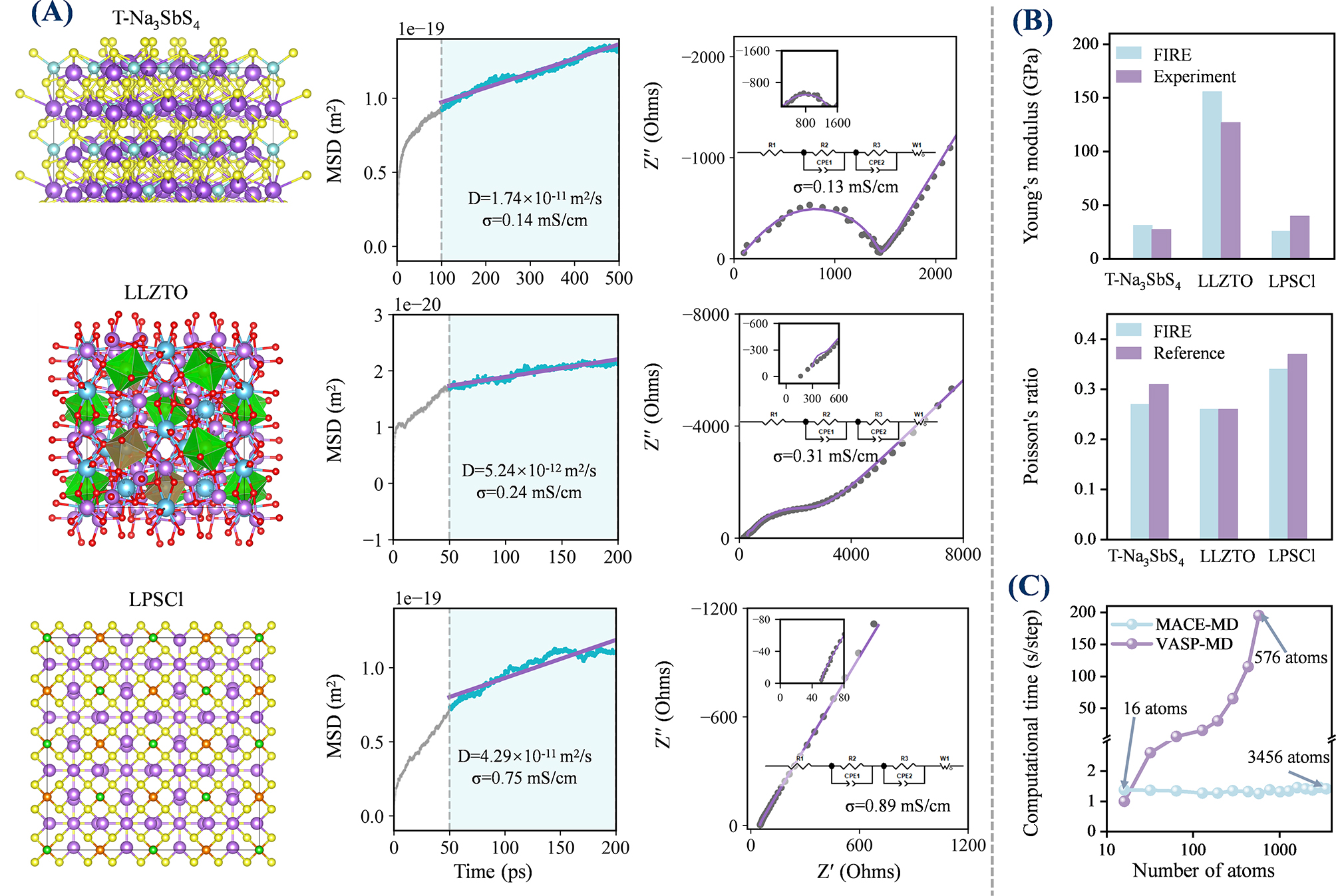}
\caption{
Property predictions from FIRE-generated fine-tuned MLIP models. (A) First column: Optimized structures of \ch{T-Na3SbS4}, LLZTO, and LPSCl.  
Second column: Mean square displacement (MSD) curves from MACE-MD simulations at relevant temperatures.  
Third column: Room-temperature impedance spectra with equivalent circuit fits.  
(B) Comparison of predicted Young's modulus and Poisson’s ratio against experimental values. Young’s modulus was obtained by spatially averaging AFM nanomechanical maps. Reference values of Poisson’s ratios are taken from: \ch{T-Na3SbS4}~\cite{Awais2024Na3SbX3Poster}, LLZTO~\cite{yuElasticPropertiesSolid2015}, and LPSCl~\cite{AEL_Chen}.  
(C) Scaling of MD wall time per simulation step. MLIP-based MD maintains nearly constant cost up to 3456 atoms, in contrast to the cubic scaling behavior of VASP-based MD.
}
\label{fig:property-overview}
\end{figure*}

We finally evaluate their predictive power for both electrochemical and mechanical properties of the task-specific MLIPs via molecular dynamics simulations. The results are shown in Fig.~\ref{fig:property-overview}. For three representative systems (\ch{T-Na3SbS4}, \ch{Li_{6.5}La3Zr_{1.5}Ta_{0.5}O12}; LLZTO, and \ch{Li6PS5Cl}; LPSCl), the optimized atomic structures are presented in the left panel of Fig.~\ref{fig:property-overview} (A), followed by room-temperature MSD curves in the middle panel. Post-processed diffusion coefficients and ionic conductivities exhibit strong agreement with literature values~\cite{sauUnlockingSecretsIdeal2024, klarbringNaVacancyDrivenPhase2024, doraiDiffusionCoefficientLithium2018, adeliBoostingSolidStateDiffusivity2019}. The right panel shows Nyquist plots from electrochemical impedance spectroscopy, where conductivities from MLIP-MD match experimental measurements~\cite{tangUltrathinSaltfreePolymerinceramic2021} closely, highlighting the transferability and reliability of our fine-tuned models trained by the FIRE approach. Based on the Nernst-Einstein relation with a Haven-ratio correction~\cite{klarbringNaVacancyDrivenPhase2024, doraiDiffusionCoefficientLithium2018, adeliBoostingSolidStateDiffusivity2019}, these findings minimize discrepancies between simulated and measured results. In addition to electrochemical properties (e.g., ionic conductivities), interfacial mechanics play a critical role in battery performance, particularly at solid-solid interfaces. As shown in Fig.~\ref{fig:property-overview} (B), by comparing with both experimental measurements (viz. atomic force microscopy shown in Fig.~S2 in SI) for Young's modulus, and first-principles calculations for Poisson's ratio, the fine-tuned MLIPs capture mechanical response with relatively high accuracy. Finally, computational efficiency is benchmarked in Fig.~\ref{fig:property-overview} (C). Dynamical simulations through fine-tuned MLIPs maintain nearly constant wall-time per MD step from 16 to 3456 atoms, whereas first-principles methods exhibit cubic scaling and are practically limited to 600 atoms due to memory bottlenecks. Together, Fig.~\ref{fig:property-overview} demonstrates that the fine-tuned models from FIRE not only reproduce transport and mechanical properties in close agreement with experiments, but also extend atomistic simulations to time and length scales inaccessible to ab initio methods. Details about sample preparation and characterization (viz. X-ray diffraction shown in Fig.~S3) are given in Section~S.4 in SI.

Finally, we illustrate how a FIRE model is used in practice, and what should be checked before deployment, with the \ch{Na/Na3SbS4} model of Fig.~\ref{fig:accuracy} (hereafter FIRE; 0.5 meV/atom and 19 meV/\AA{} on its test set; the repaired variants below are named by the data added to its fine-tuning set). Placed on a freshly built 2712-atom Na$|$\ch{Na3SbS4}$|$Na sandwich, in which bcc Na slabs cover both faces of a $2\times2\times3$ \ch{Na3SbS4} slab (Section~S.5 in SI), the model collapsed within about 50~fs at 100~K: atoms fell into spurious minima of the potential-energy surface at sub-physical separations (Na--Na 1.11~\AA, S--S 1.28~\AA, Na--S 1.36~\AA). Test-set errors are evaluated on thermal configurations and say nothing about the strongly compressed distances that a reconstructing interface samples.

\begin{figure}[H]
\centering
\includegraphics[width=16 cm]{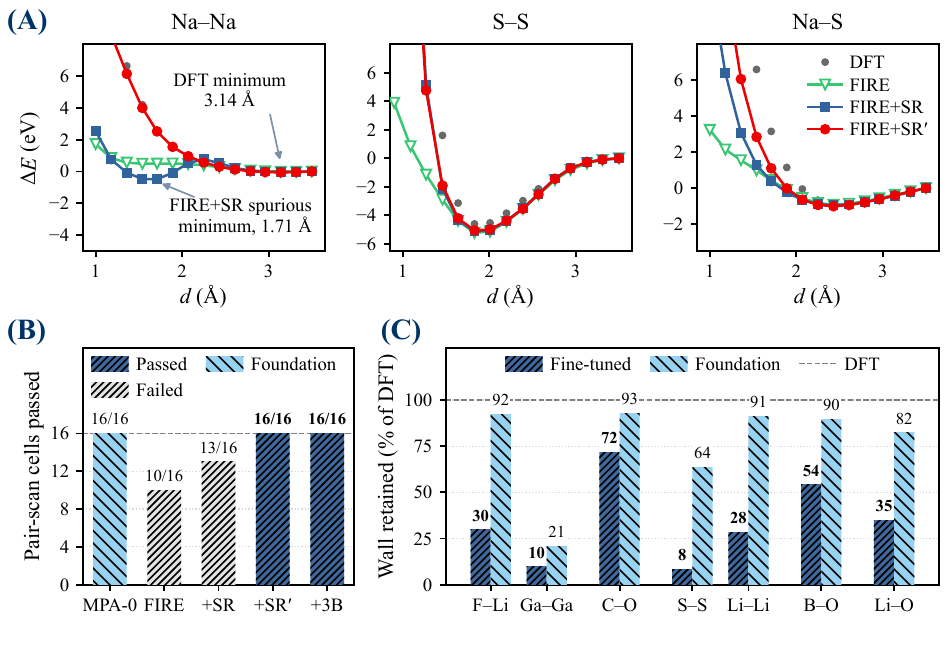}
\caption{Stress test and repair of the \ch{Na/Na3SbS4} model: (A) vacuum dimer energies of Na--Na, S--S and Na--S from DFT (grey points) and from the fine-tuned model before repair (FIRE, green open inverted triangles), after the first repair (FIRE+SR, blue squares) and after the second repair (FIRE+SR$'$, red circles); energies are relative to the value of each curve at 3.5~\AA{} (axis truncated at 8~eV), and the spurious Na--Na minimum of FIRE+SR (1.71~\AA) and the DFT minimum (3.14~\AA) are marked; (B) pair-scan cells passed (of 16) for the foundation model (MPA-0, sky-blue bars) and the four fine-tuned models (navy bars where the stress test passed, light-grey bars where it failed; +SR, +SR$'$ and +3B abbreviate FIRE+SR, FIRE+SR$'$ and FIRE+SR$'$+3B); the hole hunt (50 random starts) collapsed 3 times for FIRE+SR and never for the other models; (C) deep-compression wall height at 0.9~\AA{} (1.2~\AA{} for Li--Li), measured from the minimum of each curve, as a percentage of the DFT wall (dashed line), fine-tuned model (navy) versus foundation model (sky-blue), for F--Li and Ga--Ga in \ch{LiCl/GaF3}, C--O in \ch{Li2CO3/LiF}, S--S and Li--Li in \ch{Li/Li6PS5Cl}, B--O in \ch{Li3PS4/Li3B11O18} and Li--O in \ch{Li_{7}La_{3}Zr_{2-x}M_{x}O_{12}}. Details and absolute values are given in Section~S.5 in SI.}
\label{fig:application}
\end{figure}

The defect is easy to locate. We labelled 90 dimer configurations (six element pairs, 15 distances each from 0.9--1.4~\AA{} up to 3.5~\AA) with DFT at the settings of the training labels (Section~S.5). FIRE is markedly softer than DFT in the compressed range of all six pairs (three are shown in Fig.~\ref{fig:application}(A)): at the shortest distance sampled it retains only 7--28\% of the DFT wall height, whereas the foundation model retains 33--74\%. Fine-tuning on thermal data thus erodes the short-range repulsion inherited from pretraining, an effect recently described by Tompa et al.~\cite{Tompa2026finetuning}.

The remedy stays within the FIRE workflow: the fine-tuning set is augmented with short-range configurations and the model is refit from the foundation weights with the same replay protocol. A first repair set of 220 DFT-labelled frames (pair-squeezed cells, collapse-path snapshots, compressed cells and high-temperature frames) gave FIRE+SR, which restored the S--S and Na--S walls but over-corrected Na--Na, producing a spurious minimum at 1.71~\AA{} where DFT places it at 3.14~\AA{} (Fig.~\ref{fig:application}(A)). A rebuilt 260-frame set with 128-atom Na cells, more interface frames and \ch{Na2} dimers gave FIRE+SR$'$, which restores all six pairs and leaves the accuracy on the original validation set unchanged (0.18 meV/atom and 10 meV/\AA{} on its 214 bulk frames, identical to FIRE). FIRE+SR$'$ is the model we release; because the repair changes the potential only below about 2.2~\AA, outside the region sampled by thermal bulk and slab configurations, it is indistinguishable from FIRE on every quantity of Figs.~\ref{fig:accuracy}--\ref{fig:property-overview} and is the model used in the bulk simulations below.

Both defects are caught by an inexpensive stress test (Fig.~\ref{fig:application}(B); Section~S.5). Each pair is scanned from 0.8 to 4.0~\AA{} as a vacuum dimer and embedded in a 128-atom bulk cell and a 272-atom interface cell (16 cells), checking for spurious minima, attractive forces in the compressed range and the wall height; a hole hunt then relaxes 50 randomly perturbed 712-atom sandwiches with a capped minimizer and counts collapses to sub-physical contacts. The foundation model passes 16/16 cells with 0/50 collapses; FIRE fails six cells (10/16); FIRE+SR fails the three Na--Na cells (13/16) and collapses in 3/50 hunts; FIRE+SR$'$ and FIRE+SR$'$+3B (the same set extended by three-body patches) pass 16/16 with 0/50. The stress test requires only model evaluations and no additional DFT.

The erosion is not specific to \ch{Na/Na3SbS4}. Fig.~\ref{fig:application}(C) anchors seven pairs from five of the six systems with DFT dimer curves and shows the wall height at 0.9~\AA{} (1.2~\AA{} for Li--Li) as a percentage of the DFT value. For all seven pairs the fine-tuned models retain only 8--72\% of the DFT wall (S--S in \ch{Li6PS5Cl} is the extreme case, 10.6 versus 127.4~eV), whereas the foundation model retains 64--93\% for six of them; Ga--Ga is the one pair where the foundation model is itself soft (21\%), and fine-tuning erodes it further, to 10\%. Bulk molecular dynamics does not reveal these defects: 50~ps simulations at 300, 600 and 900~K with the models of five systems completed without a single collapse (57/57 runs, Section~S.5), although several of them carry the eroded walls of Fig.~\ref{fig:application}(C). Two prescriptions therefore follow for FIRE users: include short-range repulsion configurations in the fine-tuning set, and run the pair-scan and hole-hunt stress test before deployment.

At the 2712-atom interface, however, FIRE+SR$'$ still failed a behavioural test: all 30 production attempts at 300~K were terminated by sub-physical contact cascades whose cores reach interatomic distances of 0.08--1.0~\AA{} (Section~S.5). A freshly reconstructing interface samples configurations beyond the bulk and slab data, so the static stress test is necessary but not sufficient; closing this gap is the task of the uncertainty-aware active learning proposed in the Conclusion.


\section{Conclusion}
In summary, this work develops FIRE, a data-efficient and accurate framework for modeling solid--solid interfaces in battery materials, with a primary focus on alleviating the \emph{sampling bottleneck} in interfacial MLIPs. By coupling MD-based configuration generation with descriptor-driven dimensionality reduction and diversity-based selection, FIRE constructs compact yet representative training sets that provide broad coverage of heterogeneous interfacial environments. With substantially fewer labeled structures than conventional workflows, FIRE achieves state-of-the-art accuracy (energy errors of 0.1--1.0 meV/atom and force errors of 8--54 meV/\AA{}) across chemically diverse interfaces and enables reliable predictions of key electrochemical and mechanical properties. A deployment test on a reconstructing \ch{Na/Na3SbS4} interface shows that this accuracy is bounded by the coverage of the sampled configurations: fine-tuning erodes the short-range repulsion inherited from pretraining, a model-only stress test exposes the defect before deployment, and a few hundred short-range configurations added to the fine-tuning set restore it. Looking forward, uncertainty-aware active learning can be naturally integrated into the sampling stage to further improve coverage of rare interfacial events such as those of a reconstructing interface and extend simulations to longer time and larger length scales, accelerating predictive design of next-generation energy materials.

\section{Methods}

\subsection*{Representative sampling}
For each system, seed configurations were taken from short molecular-dynamics runs and random structure generators and perturbed by random displacements of 0.02--0.05~\AA; 20--50 of them were labelled with DFT to pre-fine-tune the foundation model. This intermediate model drove NVT molecular dynamics at 300--600~K (1~fs time step, 200--500~ps), which produced about 20{,}000 candidate configurations per system. Each candidate was described by the SOAP vectors of its 20 lowest-coordinated atoms (\texttt{dscribe}), averaged into one feature vector. The feature matrix was standardized, reduced to two principal components and clustered by $k$-means (\texttt{scikit-learn}), and one representative per cluster was retained. The selected configurations, 500--4000 per system, were labelled with DFT (Section~S.1).

\subsection*{Fine-tuning}
All models were fine-tuned from the MACE-MPA-0 foundation model (\texttt{mace-mpa-0-medium})~\cite{batatia2023foundation} with MACE 0.3.12. Replay-augmented fine-tuning mixes the task-specific data with pretraining configurations that share at least two atomic species with the target system; the replay buffer holds 1{,}000 configurations, selected by farthest-point sampling in SOAP space when more candidates are available (Section~S.1). Training used the universal loss with energy and force weights of 1 and 10 (and a stress weight of 10 for \ch{Na/Na3SbS4}), the Adam optimizer with a learning rate of $5\times10^{-4}$ and a weight decay of $10^{-6}$, a batch size of 8, an exponential moving average with decay 0.99, and 300 epochs, the last 75 with stochastic weight averaging at energy and force weights of 100 and 10. The from-scratch and random-sampling baselines of Fig.~\ref{fig:efficiency} used the same protocol; vanilla fine-tuning used energy and force weights of 10 and 1 and a batch size of 4. Training and evaluation files, their sizes and the identity of the foundation checkpoint are listed in Section~S.6.

\subsection*{Reference calculations}
Labels for \ch{Na/Na3SbS4} were computed with VASP~\cite{kresseEfficientIterativeSchemes1996} using the PBE functional~\cite{perdewGeneralizedGradientApproximation1996} and PAW potentials~\cite{blochlProjectorAugmentedwaveMethod1994}, a plane-wave cut-off of 520~eV, Gaussian smearing of 0.1~eV, an electronic convergence of $10^{-6}$~eV and a $\Gamma$-centred grid (\texttt{KSPACING} = 0.5, $\Gamma$ only for the large cells); the electronic free energy serves as the energy label. The \ch{LiCl/GaF3} labels were computed with CP2K~\cite{vandevondeleQuickstepFastAccurate2005} (PBE-D3, GTH pseudopotentials, MOLOPT basis sets, 600~Ry cut-off). The \ch{Li2CO3/LiF}, \ch{Li3PS4/Li3B11O18} and \ch{Li/Li6PS5Cl} datasets were used as published~\cite{AISSquare_Li2CO3_LiF_2025, wang2023nature, An2025LiNucleation}; the \ch{Li_{7}La_{3}Zr_{2-x}M_{x}O_{12}} dataset is described in Section~S.2. Dimer reference curves used the VASP settings above in a 16~\AA{} cubic box with spin polarization.

\subsection*{Property simulations}
Molecular dynamics was run in LAMMPS in the NVT ensemble at 298~K (Nos\'e--Hoover thermostat, 1~fs time step, 0.5~ns) for \ch{T-Na3SbS4} (1978 atoms, 3\% Na vacancies), LLZTO (1496 atoms, 5\% Li vacancies) and \ch{Li6PS5Cl} (1372 atoms, 5\% Li vacancies). Tracer diffusion coefficients were obtained from the linear regime of the mean-squared displacement, and conductivities from the Nernst--Einstein relation with Haven ratios taken from experiment (0.35, 0.40 and 0.30). Elastic constants were computed by the strain--stress method, and the moduli and Poisson's ratio from the Voigt--Reuss--Hill average (Section~S.3). Experimental procedures are described in Section~S.4.

\subsection*{Stress test}
Each element pair is scanned from 0.8 to 4.0~\AA{} in steps of 0.05~\AA{} as a vacuum dimer and, with all other atoms fixed, embedded in a 128-atom bulk cell and a 272-atom interface cell. A cell passes if the energy has no local minimum (prominence above 0.05~eV) more than 0.3~\AA{} below the physical bond length $d_{\mathrm{phys}}$, the bond-projected force is repulsive for $d<0.85\,d_{\mathrm{phys}}$, and the wall height between 1.2~\AA{} and $d_{\mathrm{phys}}$ exceeds a pair-specific threshold of 3--8~eV. The hole hunt relaxes 50 copies of a 712-atom sandwich, randomly displaced by 0.15~\AA{} and half of them compressed by 3\%, with a capped minimizer (maximum step 0.05~\AA, at most 200 steps), and counts a collapse when a pair distance falls below 2.0~\AA{} or the energy drops by more than 1~eV per atom. Both tests need model evaluations only. The criteria were fixed before the models were compared and before the production runs (Section~S.5).

\subsection*{Interface deployment}
The deployment cell is a Na$|$\ch{Na3SbS4}$|$Na sandwich of 2712 atoms ($28.66\times28.66\times100.82$~\AA$^3$). Production used the plain model without any added repulsive term: capped minimization, warm-up at 100 and 300~K, and NVT dynamics at 300~K (Langevin thermostat, 2~fs time step). A run was terminated when the temperature exceeded 2500~K or a pair distance fell below 1.35~\AA{} (Section~S.5).

\subsection*{Statistics}
Each reported error and each point of Fig.~\ref{fig:efficiency} comes from a single training run unless stated otherwise; where runs were repeated, the scatter is given in Section~S.6.

\begin{acknowledgement}

The authors acknowledge the developers of MACE and the MACE GitHub community (\url{https://github.com/ACEsuit/mace}) for their insightful feedback and implementation support. Their contributions provided the foundation for our proposed FIRE strategy. 


\end{acknowledgement}

\section{Funding}
Z. Xu and T. Zhao acknowledge the support from the National Natural Science Foundation of China (NNSFC) (grants No. 12426304 and 12325113), and the Science and Technology Commission of Shanghai Municipality (STCSM) (grant No. 23JC1402300). S.-H. Bo acknowledges the support from NNSFC (grants No. 22222204, 22393902 and 12426304), and the STCSM (grants No. 24DZ3001402 and 23TS1401600). T. Zhao acknowledges the support from the Natural Science Foundation of Chongqing (grant No. CSTB2024NSCQ-MSX1238).

%
%
%
\begin{suppinfo}

The following files are available free of charge. \par 
\begin{itemize}
	\item S.1 Details of the FIRE Framework
	\item S.2 Details on Dataset Generation
	\item S.3 Details on MD Simulations
    \item S.4 Details on Experiments
    \item S.5 Details of the Application Example
    \item S.6 Model Identity, Training Protocol and Evaluation Matrix
\end{itemize}

\end{suppinfo}

\section{Data availability}
The DFT-labelled training and evaluation sets of all six systems, the short-range repair sets, the DFT dimer reference curves and the outputs behind Fig.~\ref{fig:application} and Sections~S.5--S.6 are deposited at \url{https://huggingface.co/datasets/shiqiao123/FIRE-battery-interface-datasets}. The fine-tuned models, including the four \ch{Na/Na3SbS4} generations of Fig.~\ref{fig:application}, are deposited at \url{https://huggingface.co/shiqiao123/FIRE-battery-interface-models}. Both repositories are private during peer review, available to the editors and reviewers on request, and will be made public upon publication.

\section{Code availability}
The sampling, training and FIRE-Audit scripts, together with the analysis scripts that generate the figures and tables, are available at \url{https://github.com/ShiQiaoL/FIRE-MLIP} (private during peer review; public upon publication).

\section{Author contributions}
X. Liu performed the calculations including the generation of the dataset, carried out the interface deployment, the stress test, the short-range repair and the associated DFT calculations, and drafted the manuscript. X. Yu contributed to the experiments including sample preparation and characterization. Y. Wang developed the framework proposed in this work, finalized the workflow with optimized settings, and wrote the manuscript. Z.-T. Sun prepared dataset of LLZTO and analyzed the results. Z. Luo conducted the MD simulations and analysis. K. Zeng performed model training and validation. T. Zhao designed the workflow of this work and polished the language. S.-H. Bo and Z. Xu supervised the research. All authors contributed to the discussions and provided feedback on the manuscript.

\noindent \textbf{Correspondence} and requests for materials should be addressed to T. Zhao or S.-H. Bo or Z. Xu.

\section{Competing interests}
All authors declare no financial or non-financial competing interests.

\bibliography{ACS_MainText_Ref}

\end{document}